\documentclass[iop]{emulateapj}
\usepackage{apjfonts}
\shorttitle{EVIDENCE OF RECONNECTION FROM CORONAL FIELD EXTRAPOLATIONS}
\shortauthors{LIU ET AL.}
\submitted{Received 2013 September 23; accepted 2013 October 17; published --}
\journalinfo{ApJL accepted version 10/17/2013}
\usepackage[colorlinks=true,linkcolor=blue,citecolor=blue]{hyperref}

\newcommand{\sm}{$\sim$}
\newcommand{\goes}{\textit{GOES}}
\newcommand{\kms}{km~s$^{-1}$}
\newcommand{\hmi}{Helioseismic and Magnetic Imager}
\newcommand{\aia}{Atmospheric Imaging Assembly}
\newcommand{\Hsi}{\textit{Reuven Ramaty High Energy Solar Spectroscopic Imager}}
\newcommand{\hsi}{\textit{RHESSI}}
\newcommand{\fermi}{\textit{Fermi}}

\newcommand{\Sdo}{\textit{Solar Dynamics Observatory}}
\newcommand{\sdo}{\textit{SDO}}

\usepackage{epsf}

\def\mathbi#1{\textbf{\em #1}}

\begin{document}
\title{Evidence for Solar Tether-cutting Magnetic Reconnection from\\Coronal Field Extrapolations}

\author{Chang Liu\altaffilmark{1}, Na Deng\altaffilmark{1}, Jeongwoo Lee\altaffilmark{1,2}, Thomas Wiegelmann\altaffilmark{3}, Ronald L. Moore\altaffilmark{4}, and Haimin Wang\altaffilmark{1}}
\affil{$^1$~Space Weather Research Laboratory, Center for Solar-Terrestrial Research, New Jersey Institute of Technology,\\University Heights, Newark, NJ 07102-1982, USA; chang.liu@njit.edu}
\affil{$^2$~School of Space Research, Kyung Hee University, Yongin 446-701, Korea}
\affil{$^3$~Max-Planck-Institut f{\"u}r Sonnensystemforschung, Max-Planck-Strasse 2, 37191 Katlenburg-Lindau, Germany}
\affil{$^4$~Heliophysics and Planetary Science Office, ZP13, Marshall Space Flight Center, Huntsville, AL 35812-9900, USA}

\begin{abstract}
Magnetic reconnection is one of the primary mechanisms for triggering solar eruptive events, but direct observation of its rapid process has been of challenge. In this Letter we present, using a nonlinear force-free field (NLFFF) extrapolation technique, a visualization of field line connectivity changes resulting from tether-cutting reconnection over about 30 minutes during the 2011 February 13 M6.6 flare in NOAA AR 11158. Evidence for the tether-cutting reconnection was first collected through multiwavelength observations and then by the analysis of the field lines traced from positions of four conspicuous flare 1700~\AA\ footpoints observed at the event onset. Right before the flare, the four footpoints are located very close to the regions of local maxima of magnetic twist index. Especially, the field lines from the inner two footpoints form two strongly twisted flux bundles (up to \sm1.2 turns), which shear past each other and reach out close to the outer two footpoints, respectively. Immediately after the flare, the twist index of regions around the footpoints greatly diminish and the above field lines become low lying and less twisted (\sm0.6 turns), overarched by loops linking the later formed two flare ribbons. About 10\% of the flux (\sm$3 \times 10^{19}$~Mx) from the inner footpoints has undergone a footpoint exchange. This portion of flux originates from the edge regions of the inner footpoints that are brightened first. These rapid changes of magnetic field connectivity inferred from the NLFFF extrapolation are consistent with the tether-cutting magnetic reconnection model. 

\end{abstract}

\keywords{Sun: activity --- Sun: flares --- Sun: magnetic fields --- Sun: X-rays, gamma rays}

\section{INTRODUCTION}\label{sect1}
Magnetic reconnection alters the magnetic field connectivity to release the accumulated free energy, which is believed to power flares and coronal mass ejections \citep[CMEs;][]{priest00}. One of the most widely known flare/CME models is the tether-cutting reconnection \citep{moore80,moore01}. According to this model, the event onset is due to reconnection of two inner legs of a sigmoidal magnetic structure, which produces low-lying shorter loops across the magnetic polarity inversion line (PIL) and longer twisted loops linking the two far ends of the sigmoid. The second stage begins when the twisted loops become unstable and erupt outward, distending the overarching envelope field. The opened legs of the envelope field subsequently reconnect back to form an arcade structure with bright ribbons at their footpoints, and the ejected flux rope escapes as a CME. This scenario is hence characterized by both a ``sigmoid-to-arcade'' and a corresponding ``four-footpoints-to-two-ribbons'' evolution \citep{liu07a,chandra11}. Many features of the tether-cutting reconnection model have been corroborated by recent advanced observations \citep[e.g.,][]{liu07b,liur10} and also simulations \citep[e.g.,][]{aulanier10}. It is notable that the first stage reconnection creates new short loops, which represents a dramatic change of magnetic connectivity close to the surface. These loops are typically not well observed, as very often they are overwhelmed by the higher post-flare arcades brightened during the second stage. Nevertheless, it was argued that the rapid and persistent enhancement of photospheric horizontal field at the flare core region could be a manifestation \citep{wang10,liu12,wangs12}.

There have been some attempts to infer the connectivity and structure of flaring magnetic field using the coronal field extrapolation and images of flare emissions. Under the nonlinear force-free field (NLFFF) assumption, \citet{liu10} traced field lines at a pre-flare time from the positions of hard X-ray (HXR) footpoints throughout the 2005 January 15 X2.6 flare. It was revealed that the magnetic reconnection proceeded along the PIL toward the regions of weaker magnetic confinement. Using the NLFFF lines traced from flare ribbons, \citet{inoue11} showed the evolution of magnetic twist around the 2006 December 13 X3.4 flare. \citet{inoue13} further explored the relationship between the spatiotemporal characteristics of twisted field structure and the occurrence conditions of different types of flares.

\begin{figure*}
\epsscale{1}
\plotone{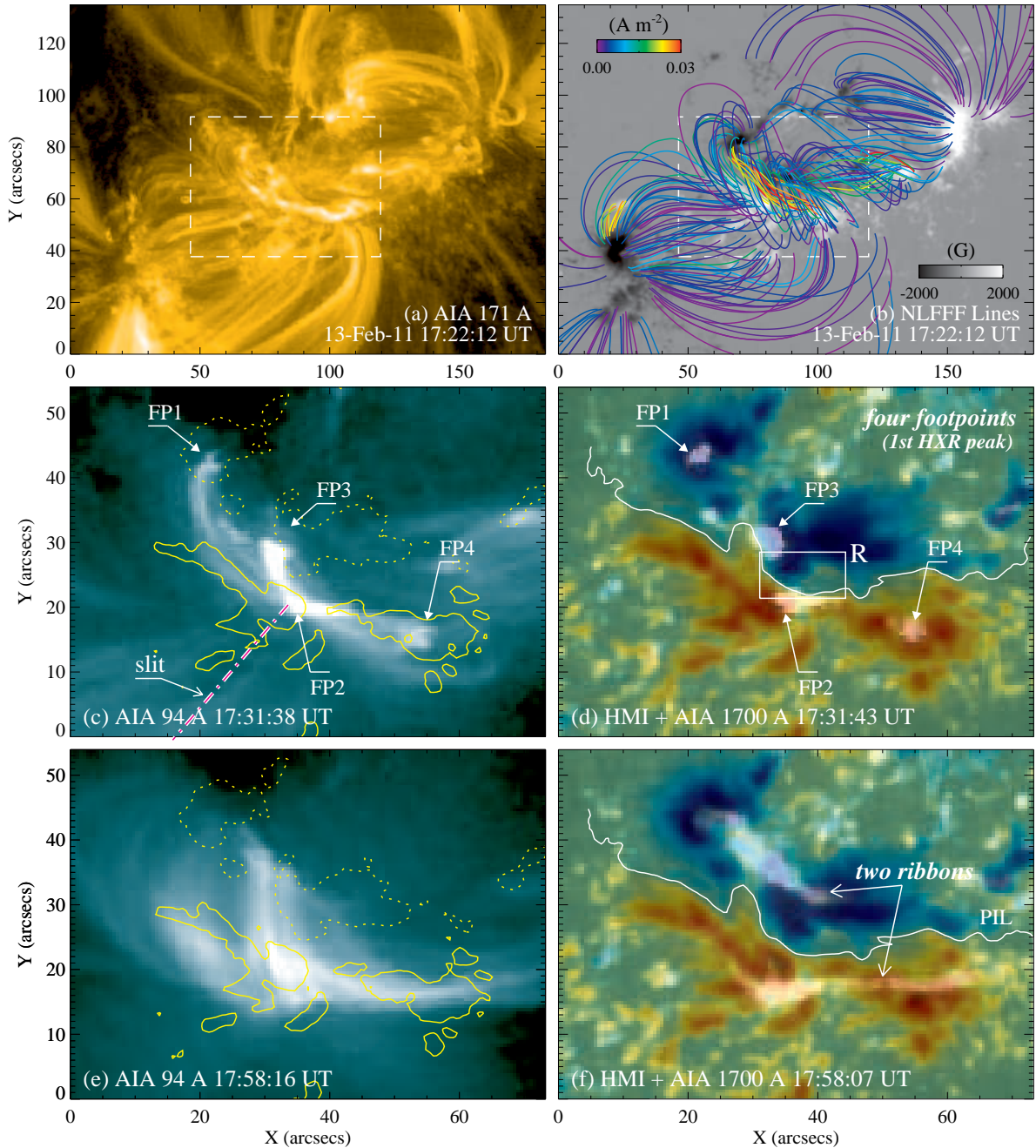}
\caption{(a--b) A 171~\AA\ image in comparison with selected NLFFF lines colored according to the vertical current density at surface. The box denotes the field of view of (c)--(f). (c and e) 94~\AA\ images overplotted with vertical field contours at levels of $\pm$1000~G. The indicated slit is used for Figure~\ref{f2}(b). (d and f) Vertical fields in blue-red scale superposed with the cotemporal 1700~\AA\ images. The box R indicates the region where horizontal photospheric magnetic field exhibits a persistent enhancement after the flare.\\ \label{f1}}
\end{figure*}

We note that a rapid change of field line connectivity during flares has not yet been clearly demonstrated using extrapolation models although it is highly desirable \citep{sun12}. In this Letter, we investigate the changes of connectivity and magnetic twist of field lines closely associated with an M6.6 flare on 2011 February 13, observations of which suggest the occurrence of the tether-cutting reconnection. For this goal, we utilize NLFFF extrapolations to trace field lines from the positions of the initial flare footpoints, at times immediately before and after the flare using the 12 minute cadence vector magnetograms from the \hmi\ \citep[HMI;][]{schou12} on board the \Sdo\ \cite[\sdo;][]{pesnell12}.

\begin{figure}
\epsscale{1.17}
\plotone{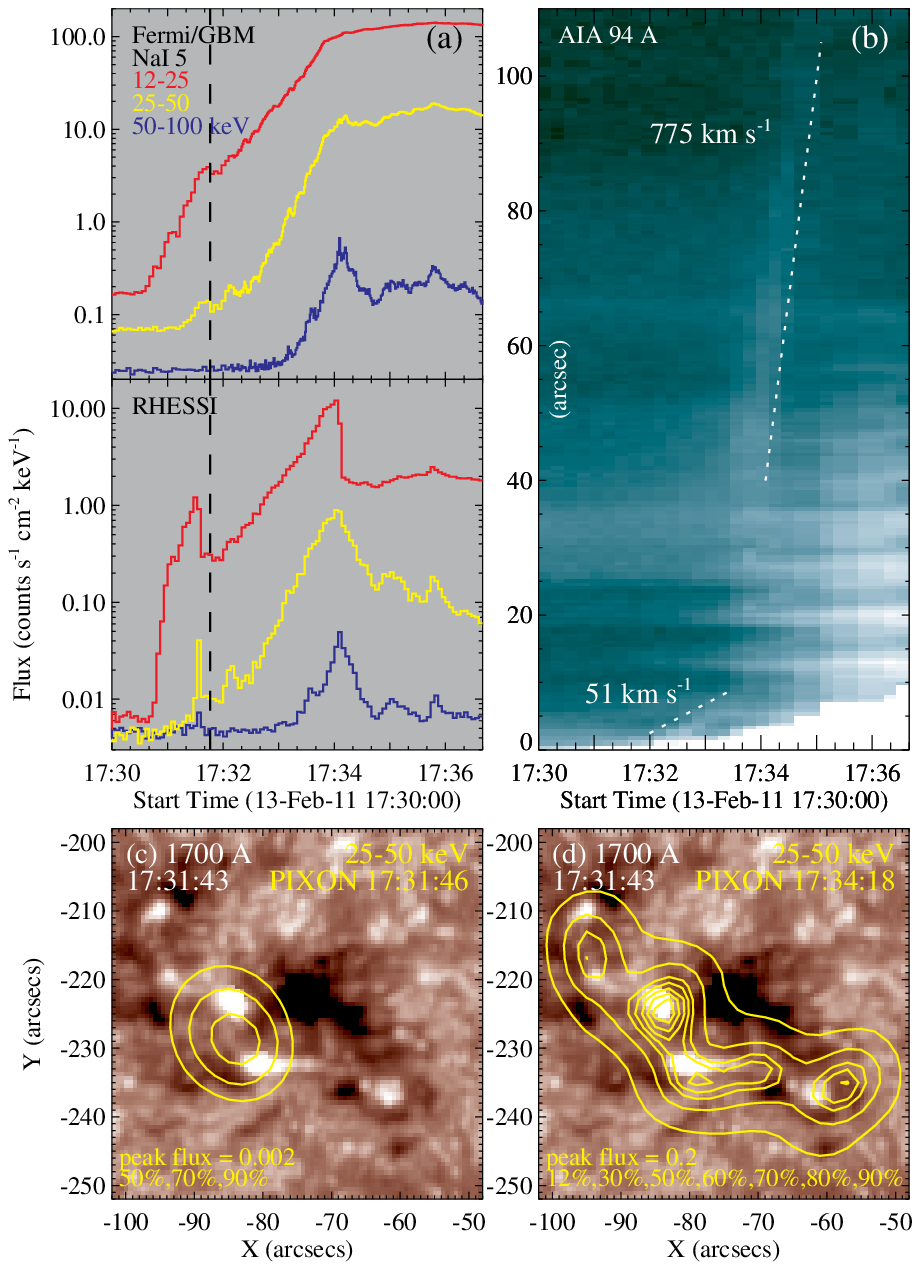}
\caption{(a) X-ray light curves. \hsi\ switched attenuators at 17:31:36 and 17:34:08~UT, which causes spurious jumps. (b) Time slice using 94~\AA\ images along the slit direction shown in Figure~\ref{f1}(c). (c--d) \hsi\ HXR images, with a unit of photons~cm$^{-2}$~s$^{-1}$~arcsec$^{-2}$. \\ \label{f2}}
\end{figure}

\section{OBSERVATIONS AND DATA PROCESSING}
The HMI vector magnetic field data are derived using the Very Fast Inversion of the Stokes Vector algorithm \citep{borrero11}. The 180$^{\circ}$ azimuthal ambiguity is resolved with the minimum energy method \citep{metcalf94,leka09a}. We used the latest version in Space weather HMI Active Region Patches (SHARP) \citep{turmon10}, and chose the format remapped using Lambert equal area projection. The observed fields were further transformed to heliographic coordinates \citep{gary_hagyard90}. After preprocessing the photospheric boundary to best suit the force-free condition \citep{wiegelmann06}, we constructed NLFFF models using the ``weighted optimization'' method \citep{wiegelmann04}. In considering measurement errors in photospheric vector magnetograms (especially the transverse field $B_T$) \citep{wiegelmann10}, we set the Lagrangian multiplier $v=0.001$ that controls deviations between the model field and the observed field and the mask function $\propto B_T$ that controls the error incorporation \citep{wiegelmann12}. The calculation was performed using 2~$\times$~2 rebinned magnetograms within a box of 372~$\times$~188~$\times$~256 uniform grid points, which corresponds to \sm267~$\times$~135~$\times$~184~Mm$^3$ of balanced magnetic fluxes. In Figure~\ref{f1}(b), we present the NLFFF model at 17:22:12~UT right before the 2011 February 13 M6.6 flare. Qualitatively, the selected extrapolated field lines in the whole active region show a reasonable agreement with the plasma loops seen in a 171~\AA\ image (Figure~\ref{f1}(a)) taken by the \aia~(AIA; \citealt{lemen11}) on board \sdo. A quantitative comparison, however, is out of the scope of this study (but see \citealt{wiegelmann12}).

To identify reconnection-related footpoints (ribbons) and hot coronal loops, we used AIA 1700~\AA\ continuum (5000~K) and 94~\AA\ (Fe~{\sc xviii}; 6.3~MK) images, respectively. Remapping was also performed for these images in order to align them with the vector magnetograms. Flare HXR emission was registered by the \Hsi\ \citep[\hsi;][]{lin02} and also \fermi\ Gamma-Ray Burst Monitor \citep[GBM;][]{meegan09}. \hsi\ PIXON images \citep{hurford02} in the nonthermal energy range (25--50~keV) were reconstructed using detectors 3--8 with 12~s integration time.

\section{OBSERVATIONAL EVIDENCE FOR TETHER-CUTTING}
The 2011 February 13 M6.6 flare started at 17:28~UT, peaked at 17:38~UT, and ended at 17:47~UT in \goes\ 1--8~\AA\ flux. To aid the interpretation of field modeling results, we first describe the most pronounced event features.

First, cotemporal with the initial small HXR peak up to 50~keV that is clearly detected by both \fermi\ and \hsi\ at 17:31:43~UT (the dashed line in Figure~\ref{f2}(a)), four flare footpoints FP1--FP4 are conspicuously brightened in 1700~\AA, with the PIL running between FP2 and FP3 (Figure~\ref{f1}(d)). Right after the flare at 17:58:07~UT, the footpoints have evolved to two elongated flare ribbons on either side of the PIL (Figure~\ref{f1}(f)).

Second, at 17:31:38~UT, two groups of hot loops connecting FP1--FP2 and FP3--FP4 as well as brighter and shorter loops FP2--FP3 can be clearly seen in 94~\AA\ (Figure~\ref{f1}(c)). At the same time, the centroid of the \hsi\ 25--50~keV image is located between FP2--FP3 above the PIL (see Figure~\ref{f2}(c)), suggesting a possible coronal HXR source produced by the reconnection between the above two loop systems. Subsequently, longer loops apparently rooted at FP1 and FP4 begin to erupt outward. The ejecting speed (in the image plane) accelerates from 51~\kms\ to 775~\kms\ around the HXR peak at \sm17:34~UT \citep{temmer08,liur10}, as measured with the time slice image (Figure~\ref{f2}(b)) along a slit drawn in Figure~\ref{f1}(c). These erupting loops are most probably responsible for the observed type II radio burst \citep{cho13} and partial halo CME \citep{yashiro12} associated with this flare. Around the HXR peak, four HXR footpoint-like sources cospatial with FP1--FP4 are imaged (see Figure~\ref{f2}(d)), which corroborates the reconnection scenario above. Moreover, behind the upward eruption, arcade fields have formed at 17:58:16~UT (Figure~\ref{f1}(e)) joining the two ribbons.

Third, at the region R sandwiched between FP2 and FP3 (see Figure~\ref{f1}(d)), photospheric horizontal field exhibits a rapid (within 30 minutes) and significant (28\%) enhancement, which is clearly related to the flare \citep{liu12}. This change is also persistent implying a permanent restructuring of near-surface field, which would be consistent with the enhanced connectivity between FP2 and FP3 \citep[also see][]{wangs12}.

Altogether, these characteristics of event morphology and dynamics are in line with those afore-introduced \citep{moore01,liu07a,liu07b,liur10,liu12} hence strongly indicate the occurrence of tether-cutting reconnection between loops FP1--FP2 and FP3--FP4, which may be triggered by either the converging photospheric flows \citep{liu12} or the evolution of small-scale magnetic patches \citep{kusano12,toriumi13}.

\begin{figure*}
\epsscale{1}
\plotone{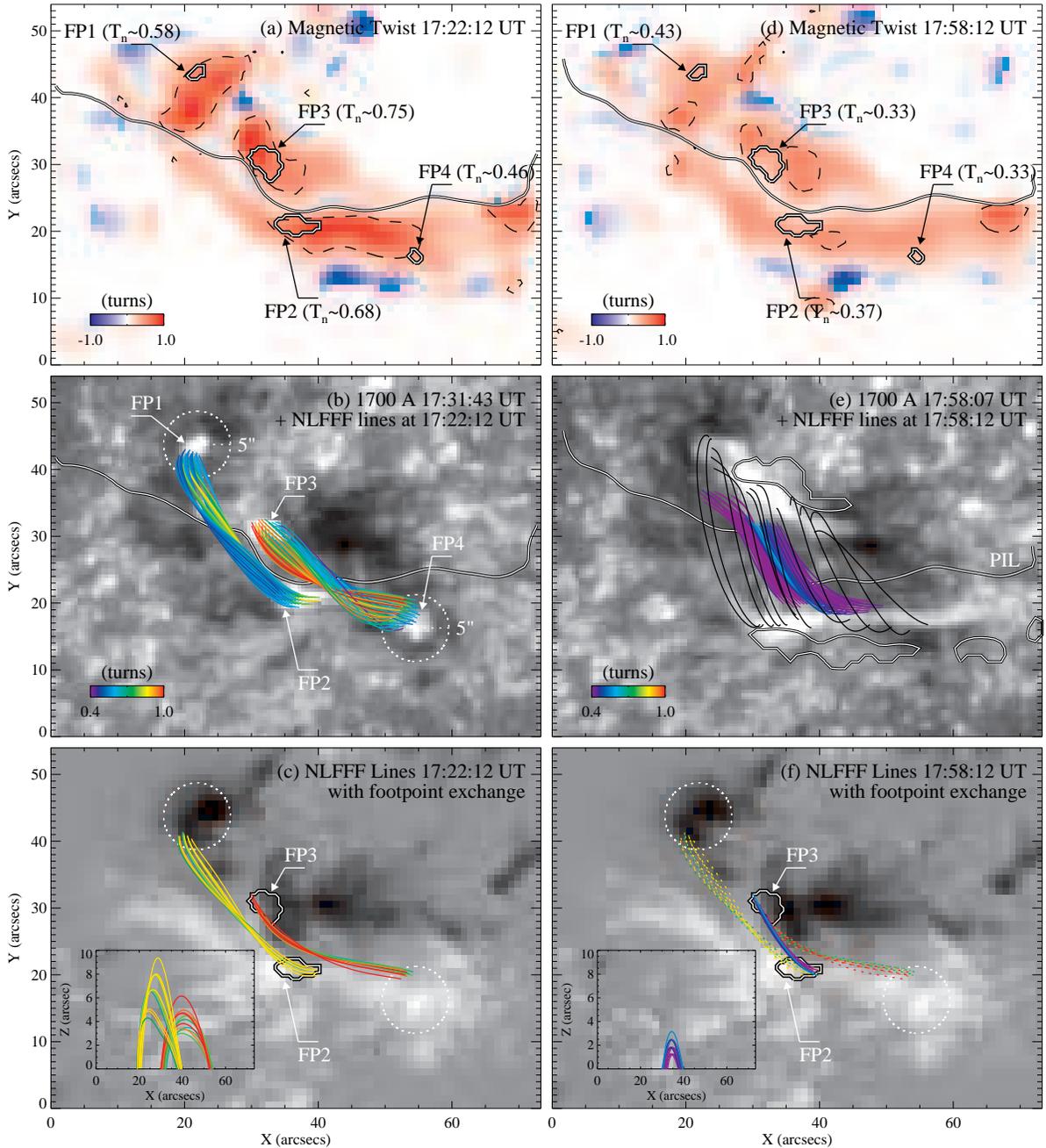}
\caption{(a and d) $T_n$ maps superimposed with dashed contours of 0.5 turns and white contours outlining the footpoints FP1--FP4. (b and e) NLFFF lines traced from FP2 and FP3 and colored according to $T_n$ at their footpoints. In (e), the black lines connecting the ribbons are not color coded, and the contours depict the regions of negative $T_n$ in (d). (c and f) Field lines exhibiting a footpoint exchange. The insets are the side view from the south.\\ \label{f3}}
\end{figure*}

\section{MAGNETIC RESTRUCTURING DEMONSTRATED FROM\\NLFFF MODELS}
We use two parameters to quantify the nonpotentiality of the force-free magnetic configuration. The force-free parameter $\alpha$ represents the degree of twist corresponding to the helicity generated by the electric current parallel to the field line \citep[e.g.,][]{torok10}. To compute the distribution of $\alpha$ on the photosphere, we use the preprocessed magnetograms and the vertical ($z$) component of $\nabla \times \mathbi{B} = \alpha \mathbi{B}$ \citep[e.g.,][]{sun12}:

\begin{equation}
\alpha = \mu_0 \frac{J_z}{B_z} = \frac{1}{B_z} \left( \frac{\partial B_y}{\partial x} - \frac{\partial B_x}{\partial y} \right) \ .
\end{equation}

\noindent Following \citet{inoue11}, the twist (in number of turns) of a force-free field line is then given by the twist index:

\begin{equation}
T_n = \frac{1}{4\pi} \int \alpha dl = \frac{1}{4\pi} \alpha L \ ,
\end{equation}

\noindent where $L$ is the field line length. Note that we calculate $\alpha$ only at pixels with $B_z \geq 30$~G and measure $T_n$ only for closed field lines in the flaring region. As the force-free condition is not perfectly guaranteed in the model, $\alpha$ at the two feet of a field line may not be equal, with a typical difference relative to their mean value ($\overline{\alpha}$) of \sm20\% for field lines from the regions of FP1--FP4. We thus take $\overline{\alpha}$ as an approximation in calculating $T_n$ \citep{inoue11,inoue13}. We also avoid flare-time magnetograms and make a comparative study using the available data right before (17:22:12~UT) and after (17:58:12~UT) the flare.

We draw in Figure~\ref{f3}(a) the pre-flare twist index map at 17:22:12~UT, which shows that most areas possess a positive twist, consistent with the positive sign of the magnetic helicity in this active region \citep{liuyang12}. The black dashed contours surround the regions of $T_n \geq 0.5$, where twisted lines with more than half-turn twist occupy. These strongly twisted locations are similar to those derived using a different NLFFF extrapolation method for an earlier time at 16:00~UT \citep{inoue13}. Interestingly, the initial four flare footpoints FP1--FP4 (white contours) lie very close to the local maxima of the strongly twisted regions. The mean $T_n$ of the inner two footpoints FP2 and FP3 are 0.68 and 0.75, respectively, while those of the outer footpoints FP1 and FP4 are smaller. As we defined all footpoint regions in terms of a fixed 1700~\AA\ intensity, the areas of the weaker FP1 and FP4 might be underestimated (cf. Figures~\ref{f1}(d) and \ref{f3}(a)).

\begin{figure}
\epsscale{1.17}
\plotone{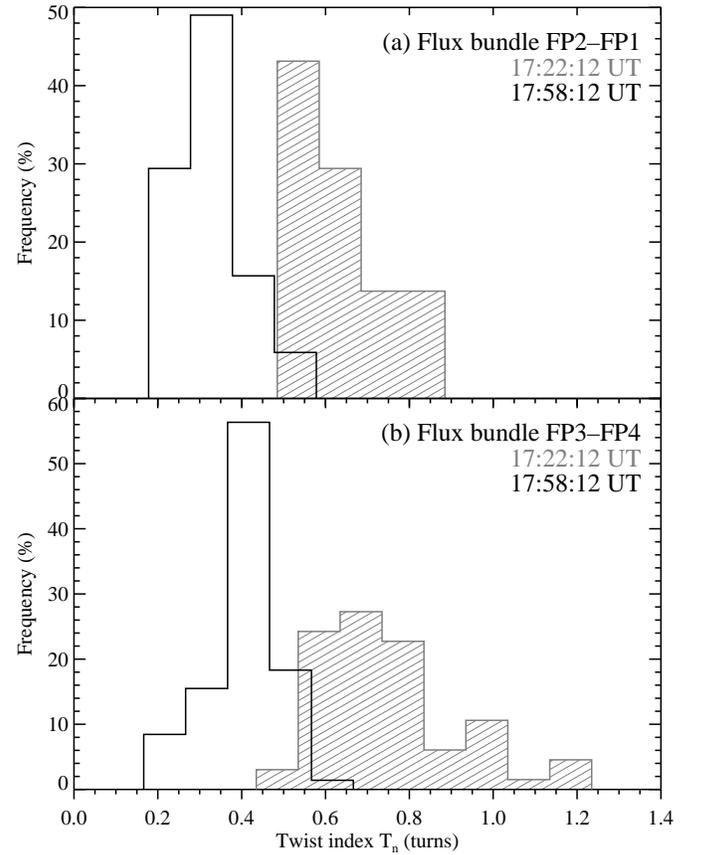}
\caption{$T_n$ histograms of flux bundles right before and after the flare.\\ \label{f4}}
\end{figure}

Figure~\ref{f3}(b) shows the magnetic field lines traced from FP2 and FP3 that are colored according to $T_n$. We only plot 51 and 66 field lines from FP2 and FP3 that end very close to FP1 and FP4, respectively, within a 5\arcsec\ radius from their centers (denoted by the dotted circles in Figure~\ref{f3}). These field lines make up a significant portion (76\% and 88\%) of all field lines stemming from the defined integer pixels of FP2 and FP3. Histograms of their $T_n$ are also plotted in Figure~\ref{f4}. It is obvious that these fields form two elongated flux bundles, which shear past each other. Almost no field lines, however, directly link FP2 and FP3 at this pre-flare time (17:22:12~UT). The model displays an overall similarity to the coronal field configuration recognized in the corresponding 94~\AA\ image (Figure~\ref{f1}(c)). Both flux bundles possess a twist index $\gtrsim$0.5 turns. FP3--FP4 bears an overall stronger twist up to 1.2 turns, and the most twisted field lines (red) are positioned closer to FP2--FP1 and the centroid of the coronal HXR source.

Since our main goal is to explore the flare-associated magnetic restructuring using NLFFF models, we repeat our analysis for vector data at 17:58:12~UT immediately following the flare. Particularly, we plot in Figure~\ref{f3}(e) the extrapolated field lines from the same starting points as those in Figure~\ref{f3}(b). Several noteworthy characteristics are described below.

First, the strongly twisted regions with $T_n \geq 0.5$ are greatly diminished (Figure~\ref{f3}(d)), suggesting the dissipation of the accumulated helicity in the flare volume. The mean twist index of FP2/FP3 regions have decreased for about 50\% on the surface, and that of both the flux bundles have weakened to below \sm0.6 turns (Figure~\ref{f4}).

Second, the modeled magnetic connectivity shows a dramatic change, with field lines in a different orientation now apparently overarching FP2 and FP3. Naturally, we intend to identify reconnection between loops FP2--FP1 and FP3--FP4, which would involve a footpoint exchange to produce a new loop FP2--FP3. It is found that among the 66 field lines at 17:58:12~UT that start from FP3, 11 of them end at points within the FP2 region. These field lines have a total flux of \sm$3 \times 10^{19}$~Mx, about 11\% of the total flux from the FP3 region. If we then perform field line tracing using the 17:22:12~UT magnetogram and start from their ending positions in the FP2 region, all the resulted field lines reach close to FP1 within 5\arcsec. Hence these field lines have underwent the footpoint exchange and are plotted in Figures~\ref{f3}(c) and (f) to demonstrate the reconnection. It can be clearly seen that before the flare at 17:22:12~UT, these field lines are highly sheared passing each other; then after the flare at 17:58:12~UT, their inner footpoints FP2/FP3 become connected with less twisted loops (also see the insets of side views). It is particularly remarkable that the highest twisted fields (red lines) among these loops are anchored at the eastern edge area of the FP3 region. By comparing the present 1700~\AA\ image at 17:31:43~UT with the previous two frames at 17:31:19~UT and 17:30:55~UT, it turns out that these areas and also the western edge area of the FP2 region are in fact brightened first. Therefore, these highly sheared loops with the highest twist index $T_n \approx 1.1$ are among the first to reconnect in the flare. For other post-flare field lines that originate from FP3 but do not land at the FP2 region (see Figure~\ref{f3}(e)), the mean deviation from FP2 is \sm4\arcsec\ to the west. Similarly, 11 out of 51 field lines of FP2--FP1 rooted at the western edge of FP2 have exchanged footpoints with field lines of FP3--FP4. This reconnected flux of \sm$3 \times 10^{19}$~Mx corresponds to 13\% of the total flux from the FP2 region. The mean deviation of the remaining field lines is \sm5\arcsec\ to the northeast of FP3. Overall, these NLFFF modeling results directly support and demonstrate the tether-cutting reconnection between loops FP2--FP1 and FP3--FP4, which results in shorter, less twisted loops FP2--FP3. The reconnection must also have produced longer and more twisted loops FP1--FP4, which erupt outward to become the CME.

Third, we also extrapolate field lines (black curves in Figure~\ref{f3}(e)) from several positions within the double flare ribbons at 17:58:07~UT. These field lines are weakly twisted at $T_n \approx 0.3$ and straddle the inner field lines FP2--FP3. They resemble the observed post-flare arcade fields in the cotemporal 94~\AA\ image (cf. Figures~\ref{f3}(e) and \ref{f1}(e)). These loops from the model also conform well to the tether-cutting reconnection, and correspond to the reconnected envelope fields after being stretched open by the erupting loops FP1--FP4. It is also worthwhile mentioning that the two ribbons stopped the separation motion before they reached the regions with negative twist (the white contours in Figure~\ref{f3}(e)).

\begin{figure}
\epsscale{1.17}
\plotone{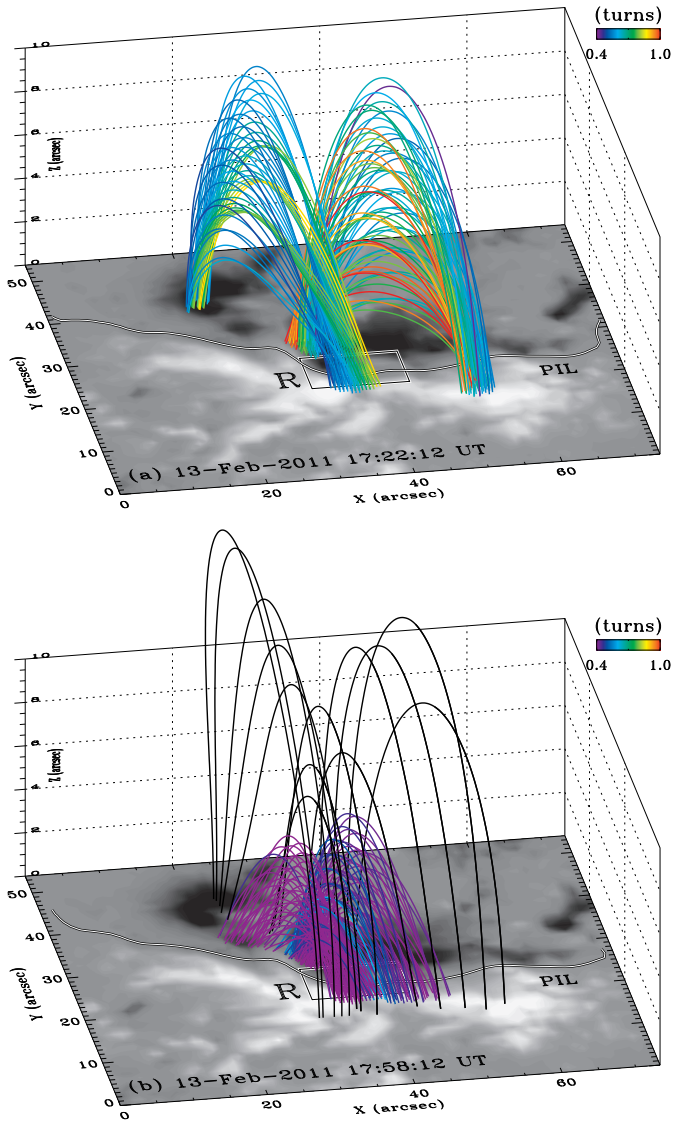}
\caption{Same as Figures~\ref{f3}(b) and (e) but drawn in 3D. The region R is the same as that plotted in Figure~\ref{f1}(d). \\ \label{f5}}
\end{figure}

Fourth, for a perspective view of the field restructuring, we re-plot Figures~\ref{f3}(b) and (e) in three-dimension (3D) in Figure~\ref{f5}. Before the flare, the two flux bundles cross each other at low heights ranging from \sm2--5\arcsec. After the flare, the inner two footpoint regions are apparently connected, with newly formed loops at a mean height of \sm3.5\arcsec\ right over the region R of enhanced horizontal field. The loops connecting the two flare ribbons arch above at a mean height of \sm14\arcsec. These models of the pre- and post-flare states are highly reminiscent of the cartoon pictures of the tether-cutting reconnection (see Figure~1 of \citealt{moore01}).

\section{SUMMARY AND DISCUSSION}
We have presented a detailed study of magnetic restructuring in the 2011 February 13 M6.6 flare utilizing both multiwavelength observations and NLFFF modeling. By investigating the extrapolated coronal fields from the positions of flare footpoints at the event onset, we demonstrate, for the first time, that the change of the magnetic field connectivity during the impulsive phase of this flare is indeed consistent with the tether-cutting reconnection model. The main results are summarized as follows.

\begin{enumerate}

\item This moderate flare is featured with four compact footpoints FP1--FP4 in 1700~\AA\ at the first minor HXR peak, which is imaged as a coronal HXR source above FP2 and FP3. Around the main HXR peak, four HXR sources cospatial with FP1--FP4 are observed. These indicate the reconnection between the hot loops FP1--FP2 and FP3--FP4 in 94~\AA. As a result, longer loops FP1--FP4 are formed and accelerated outward, leaving behind bright arcade fields above two ribbons. The expected shorter loops FP2--FP3 appear in the form of the enhanced photospheric horizontal field \citep{liu12}. These unambiguous observations strongly suggest the tether-cutting reconnection as the mechanism for this flare \citep{moore01,liu07b}.

\item At 17:22:12~UT right before the flare, NLFFF lines traced from the inner footpoints FP2 and FP3 form two twisted (up to 1.2~turns) flux bundles, which shear past each other and reach out to the regions of FP1 and FP4, respectively. Right after the flare at 17:58:12~UT, FP2 and FP3 are apparently connected with less twisted ($\lesssim$0.6 turns) shorter loops, suggesting the release of magnetic helicity. These low-lying loops (at \sm3.5\arcsec) are overlaid by higher (at \sm14\arcsec) envelope fields connecting the two flare ribbons. These model results provide a rendering of the tether-cutting reconnection picture \citep{moore01}.

\item Our NLFFF models further show that in about 30 minutes during the flare, 11 out of 66 calculated field lines of FP3--FP4 at 17:22:12~UT exchanged footpoints with field lines of FP2--FP1 to create shorter loops FP2--FP3 at 17:58:12~UT. These highly twisted and sheared field lines of FP3--FP4 are among the first to reconnect, since they stem from the eastern edge of FP3 which was brightened first. Similarly, 11 out of 51 calculated field lines of FP2--FP1 rooted at the western edge of FP2 exchanged footpoints with the field lines of FP3--FP4. The reconnected flux is estimated to be \sm$3 \times 10^{19}$~Mx, about 10\% of the flux of the FP2 (FP3) region. For other field lines that start from FP3/FP2 but do not connect to the regions of FP2/FP3 after the flare, the mean deviation from the expected positions is about 5\arcsec. These quantitative analyses further corroborate the tether-cutting reconnection and characterize its process.

\end{enumerate}

We note that subsequent 1700~\AA\ brightenings occur after \sm17:32~UT at locations between FP1 and FP3 and between FP2 and FP4, which suggests reconnection between other groups of fields not covered by our approach. Further studies combining the evolution of flare footpoint (ribbon) emissions and field modeling are promising to offer a more complete tracing of the flare reconnection process.

\acknowledgments
We thank \sdo/HMI and AIA, \hsi, and \fermi\ teams for excellent data, and the referees for valuable comments. C.L., N.D., and H.W. were supported by NASA grants NNX13AF76G, NNX13AG13G, and NNX11AO70G. J.L. was supported by the international scholarship of Kyung Hee University. T.W. was supported by DLR grant 50 OC 0904 and DFG grant WI 3211/2-1.

\end{document}